\documentclass{raa}

\usepackage{graphicx,times}
\usepackage{natbib}
\usepackage[pagebackref=true]{hyperref}

\usepackage{lineno}
\usepackage{amssymb,amsmath}
\bibpunct{(}{)}{;}{a}{}{,}

\graphicspath{{./}{figures/}}

\begin{document}
	
	\title{Half-year Evolution of a Decaying Solar Active Region and Peripheral Dimming Regions}
	
	\volnopage{ {\bf 20XX} Vol.\ {\bf X} No. {\bf XX}, 000--000}
	\setcounter{page}{1}
	
	\author{
		Jiasheng Wang\inst{1,2}, 
		Yu Xu\inst{1,*}, 
		Zhenyong Hou\inst{1}
		\footnotetext{$*$Corresponding Author, the authors contributed equally to this work.}
	}
	
	\institute{
		School of Earth and Space Sciences, Peking University, Beijing 100871, China, {\it xuyu@stu.pku.edu.cn}\\
		\and
		State Key Laboratory of Solar Activity and Space Weather, National Space Science Center, Chinese Academy of Sciences, Beijing 100190, China\\   	
	}
	
	\abstract{
		Using multi-wavelength observations from the Solar Dynamics Observatory (SDO), we investigated the six-month decay process of the solar active region NOAA AR 12738 from April to October 2019. We systematically analyzed the region's evolution by examining extreme ultraviolet (EUV) intensity variations, quantifying magnetic flux diffusion, and investigating thermodynamic changes via Differential Emission Measure (DEM) analysis. 
		This study presents the first long-term tracking of a peripheral dimming region (dark moat), revealing its continuous areal decrease over time.  
		DEM results reveal cooling plasma signatures and thermal restructuring, with the dimming region exhibiting a distinct temperature deficit in range 10$^{5.5}$ -- 10$^{5.9}$~K. 
		Potential field extrapolation identifies two dominant magnetic configurations: low-lying loops with cool plasma ($<$10$^{5.5}$ K), and high-arching structures connecting to the AR core, contributing to localized emission reduction.
		We found that the dimming is dominated by high-lying loops extending from the AR core, which are heated to temperatures above the main response of the 171~\AA\ passband ($>$ 10$^{5.8}$ K), consequently lacking plasma at the typical 10$^{5.8}$~K formation temperature. The thermal deficit, not just the absence of material, is the key driver of the reduced emission. 
		Our results demonstrate that long-duration dimming provides a valuable diagnostic for understanding active region decay, thermal evolution, and coronal magnetic restructuring.
	}
	
	\authorrunning{J. Wang et al. }            
	\titlerunning{Long Duration Coronal Dimming}  
	\maketitle
	
	\section{Introduction}\label{sect:intro}
	
	The active regions (ARs) on the surface of the Sun are the main source regions of the eruptive solar activities, such as flares, and coronal mass ejections (CMEs). The lifetime of the ARs vary from days to months \citep{2015LRSP...12....1V}. \cite{2001SoPh..199..317T} divides the evolution of the bipolar ARs into six stages. First is the emergence stage where a compact bipolar plage appears. Then, the total magnetic flux and the plage area increase rapidly, which is called the growth stage. The AR reaches its maximum development stage when its total magnetic flux and plage region are maxed, and after that, the early decay begins. The plage region starts to break up into pieces, but more than one of the sunspots remains. The AR enters the late decay stage when all the spots disappear and the plage regions fragment. The diffusion of the bipolar features continues and the remnant stage finally presents widely scattered bipolar regions which are hard to recognize.
	
	Evolution of ARs have been systematically analyzed in previous studies \citep[see][for a review]{2015LRSP...12....1V}. By analyzing the magnetic flux from the full-disk magnetograms observed by the Michelson Doppler Imager \citep[MDI;][]{1995SoPh..162..129S} onboard the Solar and Heliospheric Observatory \citep[SOHO;][]{1995SoPh..162....1D}, \cite{2008ASPC..383..397L} presented a rapid growth stage and a long-lived decay process in bipolar AR 7978. The authors also found that the distance between the weighted center of the positive/negative flux remains constant, which suggests the interaction of the AR field with the surrounding field. A local correlation tracking was conducted to show the flux cancellation and emergence processes. 
	\cite{2015ApJ...815...90U} traced the variation in the 304 \AA\ waveband during the long-term evolution of ten ARs using extreme ultraviolet (EUV) images from the Atmospheric Imaging Assembly \citep[AIA;][]{2012SoPh..275...17L} onboard the Solar Dynamics Observatory \citep[SDO;][]{2012SoPh..275....3P} and Sun Earth Connection Coronal and Heliospheric Investigation \citep[SECCHI;][]{2008SSRv..136...67H} onboard the Solar Terrestrial Relations Observatory \citep[STEREO;][]{2008SSRv..136....5K}. The corresponding magnetic flux is calculated from the Helioseismic and Magnetic Imager \citep[HMI;][]{2012SoPh..275..207S} onboard SDO. A power-law correlation between the 304 \AA\ radiance and the unsigned magnetic flux in the AR is found (i.e. radiance--magnetic flux relationships). And the Advective Flux Transport (AFT) model is used to predict the magnetic flux variation of the ARs when they rotate to the far side of the Sun, where no magnetic field observations are accessible, based on the dataset of 304~\AA\ from STEREO. 
	\cite{2017ApJ...846..165U} extend the radiance--magnetic flux relationships to the AIA 335 \AA\ waveband and the Fe XVIII 93.93 \AA\ spectral line extracted from the AIA 94 \AA\ waveband. The heating process in several decay ARs is then studied based on models. The potential field extrapolation is conducted with the boundary condition constructed from the AFT model, and the Enthalpy-based Thermal Evolution of Loops model is used to simulate the Fe XVIII and 335~\AA\ emission with a steady heating rate on each field line related to the average field strength and the line length. The radiance--magnetic flux relationship is partially reproduced, which suggests the ability of the models on estimating the radiance variance in the decay ARs.
	
	Coronal dimmings have been detected in extreme ultraviolet (EUV) bands and X-ray bands. Traditionally, coronal dimming has been interpreted as a signature of the footpoints of erupting magnetic flux ropes and the mass loss associated with coronal mass ejections (CMEs). However, high-resolution observations in recent years (especially from SDO/AIA) have revealed that coronal dimming contains essential physical information. It spans the entire evolution of a CME, from pre-eruption precursors and the main ejection phase to the subsequent coronal recovery, making it a unique diagnostic tool for studying CME triggering mechanisms, changes in magnetic connectivity, coronal mass replenishment, and interplanetary propagation. 
	\cite{2012ApJ...760L..29Z} studied the emerging dimmings (EDs) in 24 isolated ARs at their early emerging stages, calculating the dimming duration and the flux emergence rates. \cite{2021ApJ...912....1P} analyzed the emission measure (EM) in the emerging dimming region before and after the dimming began. The decrease/increase of the EM in the temperature range of lg T[5.7, 5.9]/lg T[6.2, 6.4] suggests the occurrence of heating. 
	
	Magnetic reconnection is the fundamental physical process responsible for altering magnetic connectivity \citep[see][for review]{2025LRSP...22....2V}, thereby leading to the formation, migration, and recovery of coronal dimming \citep[e.g.,][]{2024A&A...683A..15J, 2025ApJS..281...14J}. Different types of reconnection (such as flux rope--closed field reconnection or flux rope--open field reconnection) result in distinct dimming behaviors \citep[e.g.,][]{1996ApJ...468L..73M, 2007SoPh..244...25M, 2007SoPh..241..329Z, 2010AIPC.1216..440L, 2018ApJ...866...96J, 2020ApJ...903..129P}.
	When an erupting magnetic flux rope reconnects with open magnetic fields, one leg of the flux rope breaks off, and the other leg becomes rooted in the open field, producing deep dimming \citep{2006SoPh..238..117A, 2008JGRA..113.9103G, 2011ApJ...738..127L, 2021ApJ...911..118D}. This can strengthen and deepen existing dimmings and may lead to full detachment of the flux rope from the Sun. For example, \cite{2006SoPh..238..117A} noted that in the May 12, 1997 event, dimming on one side was enhanced due to reconnection with polar coronal holes (open magnetic fields). Dimming in open field regions recovers more slowly because re-establishing the coronal architecture or closing the field lines via reconnection takes time.
	
	Studies by \cite{2007SoPh..244...25M} and \cite{2007SoPh..241..329Z} showed that multiple dimming regions corresponded to different magnetic flux systems. If the erupting flux rope reconnects with magnetic fields near a null point, the footpoints of the rope may ``jump'' to distant regions, forming remote dimmings.
	CME events are often accompanied with toroidal flux change, which undergoes rapid increase followed by gradual decrease. Build-up of CME flux is highly associated with rise phase of flares/CMEs through reconnection of sheared magnetic field \citep{2020Innov...100059X}. 
	The flux measured from coronal dimmings correlates with both toroidal and poloidal flux of the Interplanetary Coronal Mass Ejection (ICME) \citep{2007ApJ...659..758Q, 2017SoPh..292...93T}.
	By measuring the magnetic flux in the dimming regions, one can estimate the total magnetic flux of the erupting rope, which can then be compared with the in situ measured magnetic flux of the ICME.
	\cite{2000JGR...10527251W} and \cite{2007ApJ...659..758Q} found that the magnetic flux in dimming regions is of the same order of magnitude as that of the ICME. 
	Distribution of vertical current in dimming regions can reflect the twist of the flux rope. For instance, \cite{2019ApJ...871...25W} detected non-neutralized vertical currents in pre-eruption dimming regions, supporting the existence of a magnetic flux rope. 
	However, direct measurements of the coronal magnetic field are still very difficult, although significant advances have been made recently \citep{2020ScChE..63.2357Y, 2020Sci...369..694Y, yang2022, 2024Sci...386...76Y, 2020Sci...367..278F, 2024SciA...10.1604S, 2025arXiv250808970C, 2025RAA....25a5010G}. Structures of the coronal magnetic fields are inferred through extrapolations of the photospheric magnetic field or by indirect methods such as dimming or current analysis.
	
	The loop length variance in the dimming region based on the Potential-field Source-surface (PFSS) model showed that the average loop length increases after the dimming began, which is in support of the magnetic reconnection heating the plasma.  
	\cite{2021ApJ...909...57S} investigated seven dimming regions around their corresponding ARs on their average temperatures and magnetic field structures. They reconstructed the magnetic fields using the PFSS model and computed the magnetic pressure distributions on the solar disk. The analysis results showed that the dimming region often owned high magnetic pressure, and they suggested that the strong magnetic fields overlying the dimming region press the coronal loops to a lower altitude where the temperature is beneath the main response temperature of AIA 171 \AA\ bands. 
	
	While coronal dimmings have been extensively studied, most existing research focuses on short-lived events lasting several hours, with little attention given to dimming regions that persist for months. Furthermore, previous work has predominantly examined temporary dimming characteristics, leaving the long-term evolutionary process of these structures largely unexplored. To address this gap, we present a comprehensive half-year study of NOAA AR 12738 and its associated long-duration dimming region, aiming to systematically characterize their coupled evolution through the AR's decay phase.
	This study is structured as follows. Section 2 provides an overview of AR 12738 and describes the observational data. Section 3 details the methodology, including DEM analysis and PFSS magnetic extrapolation. In Section 4, we quantify the temporal evolution of EUV emissions and magnetic flux diffusion, track the formation and deformation of a filament channel, and analyze the thermal and magnetic properties of both the dimming and core regions. Section 5 discusses our findings in the context of previous studies, and Section 6 summarizes the principal conclusions. Through this multifaceted analysis, we seek to establish a physical framework for understanding the formation, persistence, and eventual recovery of long-term coronal dimmings.

	\section{Observations}
	
	\begin{figure*}[ht!]
		\centering
		\includegraphics[width=\linewidth]{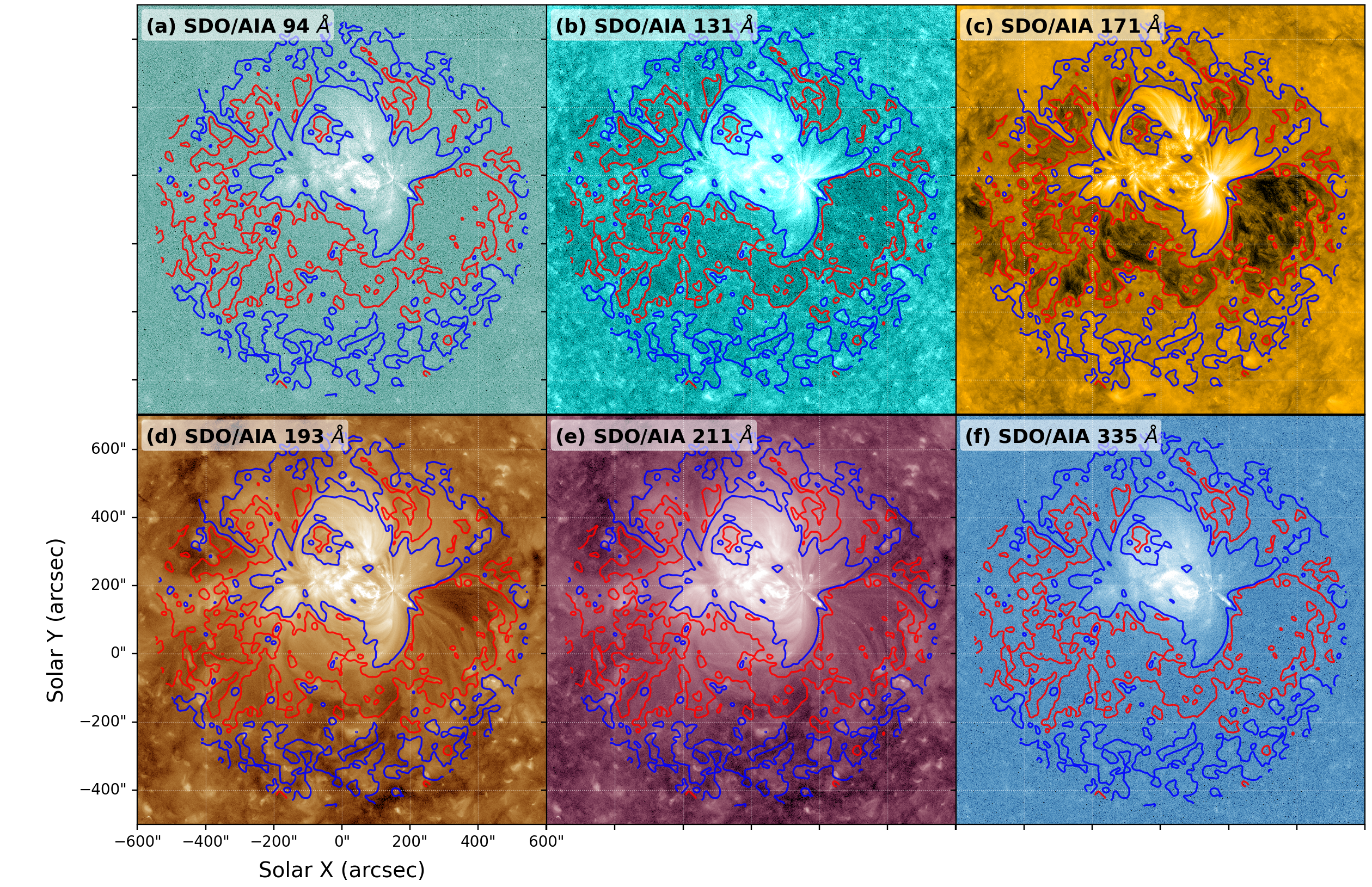}
		\caption{An overview the AR 12738 in SDO/AIA EUV channels on Apr. 14 2019. (a-f) show the active region with bright core (peripheral dimming) regions indicated by blue (red) contours in images of 94, 131, 171, 193, 211, and 335~\AA\, respectively. The contours are obtained from AIA 171~\AA\ image with method described in Section~\ref{sec:3}. }
		\label{fig:f1}
	\end{figure*}
	
	Coronal dimming signals in solar eruptive activities are usually measured with respect to EUV intensities at pre-flare time \citep{2010SoPh..262..461A, 2014ApJ...789...61M, 2018ApJ...857...62V}. In this study for the long term evolution of AR, dimming regions are defined as where intensity is below average intensity in quiet Sun of the date of observation. The description of thresholds by which different coronal features are extracted is presented in Section~\ref{sec:3}.
	The target AR 12738 first appears on the solar disk on Apr. 8, 2019 and crosses the solar central meridian on 02: 30~UT, Apr. 14, 2019 (referred as Time 1 hereafter). Figure~\ref{fig:f1} presents the bright core and peripheral dimming regions in the AR in all six AIA EUV (94, 131, 171, 193, 211, 335~\AA) channels. As there is no obvious AR on the solar disk a solar rotation before Time 1, AR 12738 completes the emergence stage (see Section 1) during the time at the far side of the sun. The moat dimming regions outside the AR are most discernible on the AIA 171 \AA\ images since the appearance of the AR, with marginal dimming signals detected in 193 and 211 \AA\ (see Figure~\ref{fig:f1}(c-f)). Another AR exists to the east beside the target AR during Time 2 to Time 4. Then a filament channel forms after Time 4 when the target AR rotates to the far side, and is observed at Time 5. The filament channel fragmented into small fibrils in the following two solar rotations and the dimming regions are gradually recovered. The filament channels and the EUV dimming are barely observable at Time 8, and the AR is considered to reach its remnant stage. Evolution of the coronal dimming in AIA 171~\AA\ is shown in Figure~\ref{fig:f2}(a1-a4, c1-c4). The dimming region resides in south of the AR, where the filament is formed in later phase. 
	
	The line-of-sight (LOS) magnetic field data is acquired from the HMI/SDO. HMI observes the photospheric magnetic field at 6173~\AA\ with a spatial resolution around 1''. It measures the polarized spectral lines split through Zeeman effect and then inverses the LOS magnetic field strength. There are two types of full-disk LOS magnetograms available with time cadences of 45 s and 720 s, respectively. In this study, we utilized the data of 720 s cadence for its higher signal-to-noise level. The uncertainty of measured field strengths is nominally below 10 Gauss. 
	Figure~\ref{fig:f2}(b1-b4, d1-d4) present evolution of magnetic field in the target AR, which shows divergence and annihilation of the bipolar field over six month period. We selected the magnetograms of the AR at particular times when its magnetic polar inversion line (PIL) crosses the solar central meridian. The crossing times of choice during the whole AR evolutionary process are listed in Table~\ref{tab1}. The Solar Rotation Numbers listed in the Table~\ref{tab1} are with respect to Carrington number 2216, and will be used in the rest of the paper to indicate time sequence instead of the times in UT formats for simplicity (e.g., Time 1 represents Apr. 14 2019 02:30 UT).
	
	\begin{table*}[ht!]
		\centering
		\begin{tabular}{cc}
			\hline
			\hline
			Carrington Number w.r.t. 2216  & Time of Central Meridian Crossing \\\hline
			1 & 2019-04-14 02:30\\
			2 & 2019-05-11 00:00\\
			3 & 2019-06-06 16:00\\
			4 & 2019-07-03 06:00\\
			5 & 2019-07-30 08:00\\
			6 & 2019-08-26 08:00\\
			7 & 2019-09-22 16:00\\
			8 & 2019-10-19 12:00\\\hline
		\end{tabular}
		\caption{Time of the central meridian crossing the Target AR}
		\label{tab1}
	\end{table*}

	\section{Data Analysis}\label{sec:3}
	
	\begin{figure*}[ht!]
		\centering
		\includegraphics[width=\linewidth]{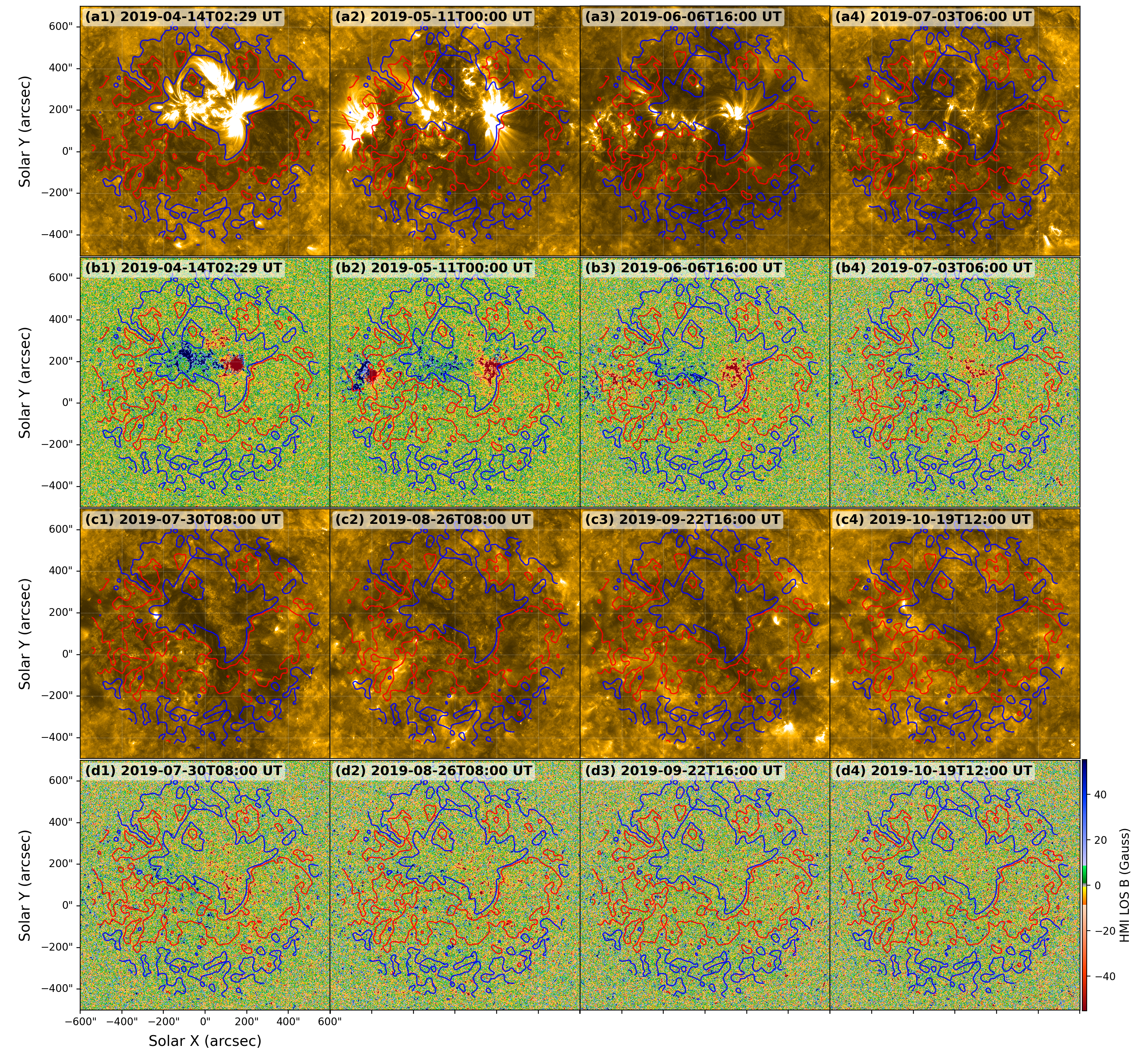}
		\caption{Evolution of the AR. (a1-a4) and (c1-c4) show the AR in AIA 171 images at 8 different central meridian crossing times. (b1-b4) and (d1-d4) show magnetograms of the AR at corresponding times. Color bar scales from -50 to 50~G. }
		\label{fig:f2}
	\end{figure*}
	
	The EUV images used are taken by the AIA on board SDO. AIA images the Sun over 1.3 solar radius in multiple wavelengths almost simultaneously. Each passband has time cadence of 12 s and spatial resolution of 1.5''. The images from six pass-bands (i.e. 94 \AA, 131 \AA, 171 \AA, 191 \AA, 211 \AA, 335 \AA) closest to the times listed in the Table~\ref{tab1} are used for analysis.  We further evaluated EUV intensity in six hour time span from the crossing times to avoid impact of transient eruptions.  
	
	The EUV images taken by AIA/SDO are used to identify different coronal structures and dimming region. The threshold value R$_\lambda$ is set to be the median of the on-disk intensity on $\lambda$ band. The dark regions in 171~\AA\ images, where the intensity is less than 0.6R$_{171}$, consists roughly of three contributions: the dimming regions, the coronal holes, and the filament channels. 
	Dimming signals in 193 \AA\ and 211 \AA\ images are defined as less than 0.75R$_{193}$ and 0.8R$_{211}$, respectively.
	The filament channel is where the intensity on the 171~\AA\ images is less than $\frac{1}{3}$R$_{171}$ , while the intensity of the coronal holes is less than $\frac{1}{3}$R$_{193}$ and $\frac{1}{3}$R$_{211}$ on 193~\AA\ and 211~\AA\ images, respectively. 
	Combining thresholds of intensity set for the three channels, the dimming regions \sout{is} are obtained after subtracting signals of filament and coronal holes from dark regions.
	The AR core is defined based on EUV images with intensity between $\frac{4}{3}$R$_{\lambda}$ and $\frac{7}{3}$R$_{\lambda}$ for $\lambda$=171, 193, and 211\AA.
	The red and blue contours in Figure~\ref{fig:f1} shows the results of the area separation method on 14 Apr. 2019.
	We determine the boundaries of dimming and core regions by applying predefined thresholds to EUV images within initially outlined areas, which are enclosed in the AR and contract over time.
	The magnetic field data is acquired from HMI onboard SDO. The line-of-sight magnetic field strength is averaged every twelve minutes by HMI for a higher signal-to-noise ratio. Figure~\ref{fig:f2} shows evolution of the AR in EUV images and magnetograms. 
	Throughout this paper, when we refer to the temperature of specific loop populations (e.g., high-lying loops heated beyond the 171 \AA\ response), this inference is based on combining two independent diagnostics: (1) PFSS extrapolation identifies the magnetic connectivity of loops in different regions; (2) DEM analysis quantifies the temperature distribution of plasma in those connected regions. We do not directly measure the temperature of individual loops, but rather infer the thermal state of magnetically connected plasma volumes.

	\subsection{DEM Calculation}\label{sec:3.1}
	
	To assess average effective temperature of different regions , differential emission meaure (DEM) is calculated, which evaluates the emissivity of a volume element in a function of temperature. For the optically thin emissions in the EUV wavelengths, the observation gives out: 
	
	\begin{equation}
		y =\int_{0}^{+\infty}\, G(T,n_e)\cdot DEM(T)\, dT
	\end{equation}
	
	where $G(T,n_e)$ is the contribution function which is sensitive to the temperature and electron density, and $DEM(T)dT = n_e^2(T)ds$ \citep{2015ApJ...807..143C}. 
	The emission measure (EM) in a temperature range of [T1,T2] is then defined as $EM(T) = \int_{T1}^{T2} DEM(T)\,dT$. We used the DEM inversion routine {\it aia\_sparse\_em\_solve.pro} \citep{2015ApJ...807..143C} to calculate the DEM(T) from the six-waveband (94  \AA, 131  \AA, 171  \AA, 193  \AA, 211  \AA, 335  \AA) EUV images observed by AIA/SDO. The DEM weighted temperature expression is adopted from Del Zanna \& Mason (2018), which is modified from calculations that was firstly introduced by \cite{2012ApJ...761...62C} as:
	
	\begin{equation}
		\log T_{DEM} = \frac{\int_{0}^{+\infty}\, DEM(\log T)\cdot\log T\, d(\log T)}{\int_{0}^{+\infty}\, DEM(\log T)\, d(\log T)}
	\end{equation}
	
	where the integration is always replaced by the series in a finite range of temperature due to sparse inversion of $DEM(T)$. 
	To estimate the uncertainties in the DEM results, we perform 500 iterations of sparse inversion. For each iteration, the observed intensity profiles are perturbed by $\delta$, which is randomly sampled from a Gaussian distribution. The standard deviation $\sigma$ of this distribution is derived from the {\it aia\_bp\_estimate\_error.pro} routine.

	\subsection{PFSS Extrapolation}\label{sec:3.2}
	
	To infer the coronal field structures in the AR core region and dimming region. We analyzed the extrapolation results of Potential-Field Source-Surface (PFSS) model, which is developed by \cite{1969SoPh....6..442S} and \cite{1969SoPh....9..131A} to give a force-free approximation of the coronal magnetic field (i.e. a potential field). In this study, the inner boundary is set on the photosphere, while the outer boundary refers to the source surface where the magnetic field is radial. The PFSS model requires a synoptic magnetogram as the input including the photospheric magnetic field of both the near-side (i.e the side facing the earth) and the far-side of the sun. The magnetic field on the solar disk has continuous observation, and the far-side information is derived from the near-side observations using the differential rotation model.
	
\begin{figure*}[ht!]
		\centering
		\includegraphics[width=0.8\linewidth]{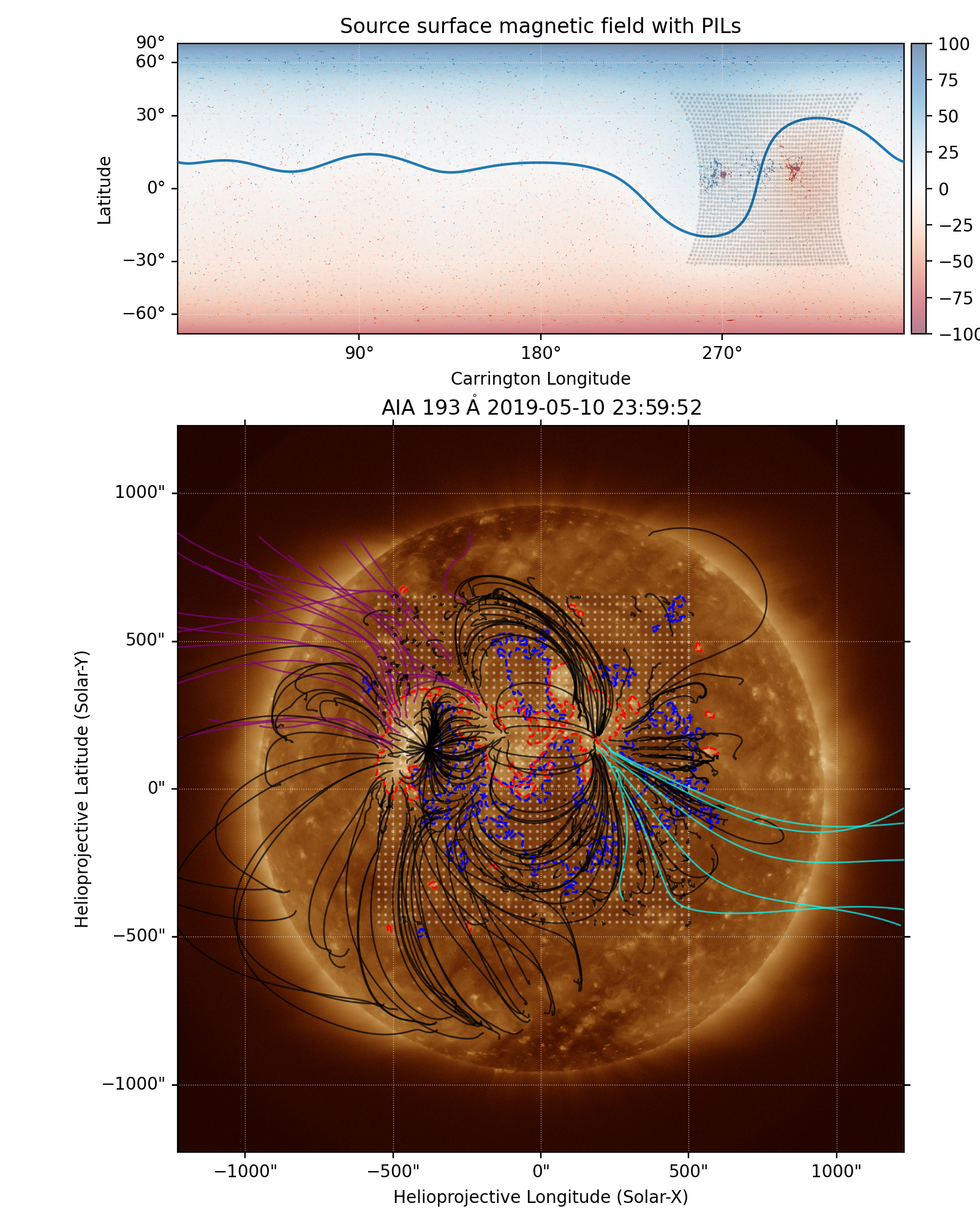}
		\caption{Synoptic magnetogram and footpoints of field tracing. The black dots in upper panel show the Carrington projected footpoints in the HMI image, each dot indicate a group of 25 footpoints around it. The blue curve indicates PIL across the AR at solar surface. The red (blue) dashed contours indicate core (dimming) region of the AR. The black and purple (cyan) streamlines represent closed and positive (negative) open field, respectively. }
		\label{fig:f3}
	\end{figure*}

	We acquired the synoptic magnetograms from HMI and conducted the PFSS model using the {\it pfsspy} package in Python \citep{2020JOSS....5.2732S}. The source surface was set to locate at R$_\mathrm{s}$ = 2.5R$_\mathrm{SUN}$ where R$_\mathrm{s}$ is the distance from the solar center and R$_\mathrm{SUN}$ is the solar radius. There were 300 calculation points along the radial direction with adapted intervals. Field lines were traced using tracer routine in {\it pfsspy} for all pixels in the magnetograms resampled with 2$\times$2 binning. And a layer of regularized grids with 10'' cell distance above the photosphere is then used as seeds to select field lines for loop height calculation (see Figure \ref{fig:f3}).
	
	\section{Results}\label{sec:4}
	
	Figure~\ref{fig:f2}(a1-a4, c1-c4) shows the evolution process of the whole AR during the eight Carrington periods on the base of AIA 171 \AA\ images. Those bright loops overlying the center part of the AR exist for the first three Carrington periods and gradually fade over time. A filament channel is formed in late July, and it then segments and disappears in the last three Carrington periods along with whole the AR. The magnetic field evolution is shown in Figure~\ref{fig:f2}(b1-b4, d1-d4). The only one sunspot shown on Apr. 14 suggests the AR has entered the early decay stage (mentioned in Section 1) since the first detection. One Carrington period later, the sunspot begins to break into pieces with the AR stepping into the late decay stage. Along with the segmentation of the filament channel, the AR gradually evolves into the remnant stage where the bipolar field is barely recognized.

	\subsection{Variation of Intensity and Thermal Energy}\label{sec:4.1}
	
	\begin{figure*}[ht!]
		\centering
		\includegraphics[width=\linewidth]{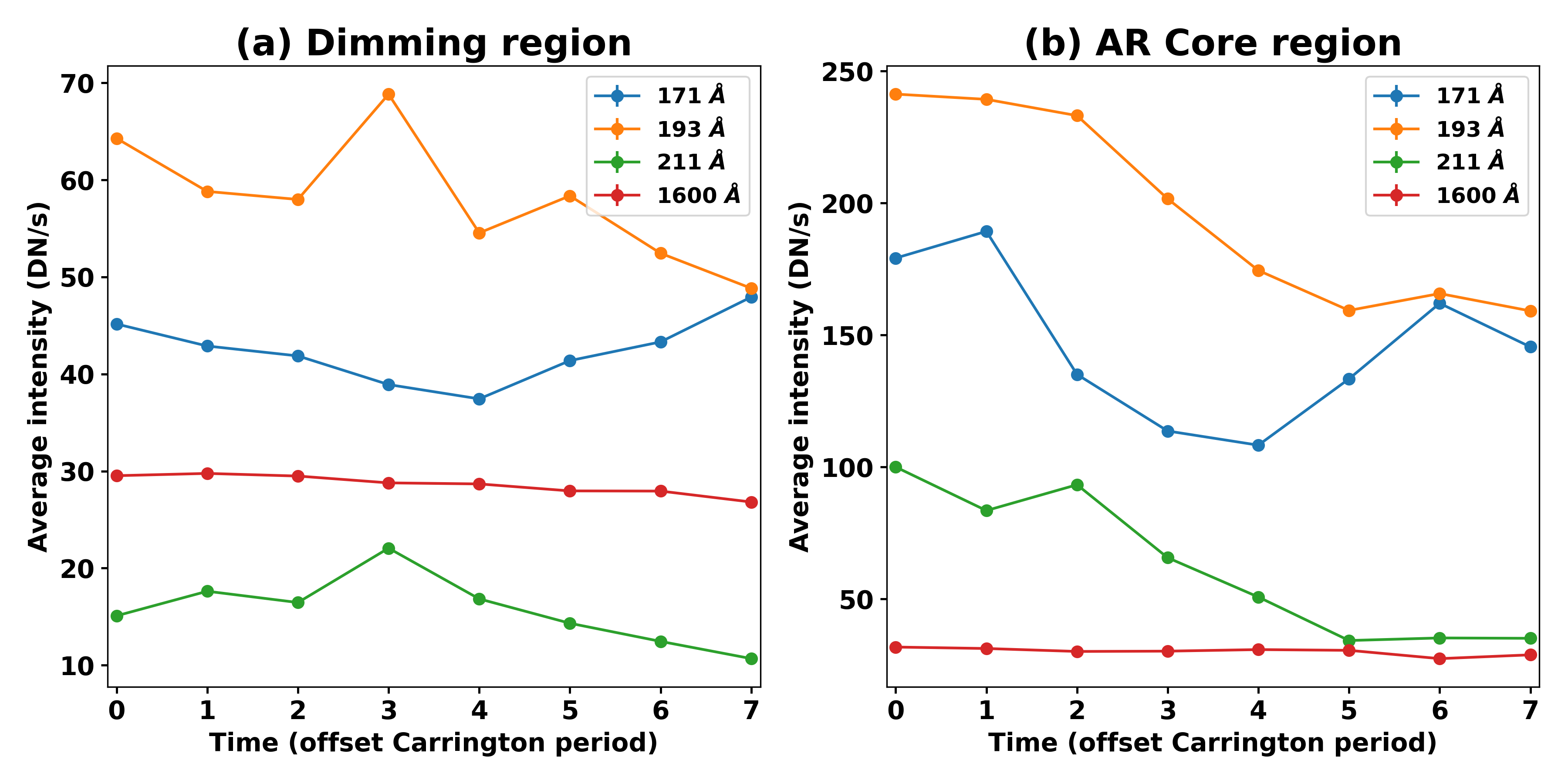}
		\caption{The average intensity variance (a) in the dimming region and (b) in the AR core region in AIA 171, 193, 211, and 1600~\AA. All the intensities are divided by the corresponding value at Time 1 (marked as 0 offset Carrington period) in Table~\ref{tab1}. Dots represent the relative intensity calculated from observations, and the curves show the varying trend.}
		\label{fig:f4}
	\end{figure*}
	
	The variation of average intensity in the AR core and the dimming regions are shown in Figure~\ref{fig:f4}. All of the intensity is divided by their very first value to show the relative variance. In the AR core region, the average intensity in both 211 \AA\ and 193 \AA\ wavebands continually decreases, which indicates the decay of the AR activity. In contrast, The intensity in the 171 \AA\ band shows an increasing tail in the last several months. The same increasing trend of the 171 \AA\ band also exists in the dimming region. The UV channel 1600 \AA\ of the upper chromosphere in the dimming region maintains a relatively steady state during the whole evolution with only a slight decrease. The 211 \AA\ band increases in the first several months and then decreases after the existence of the filament channel.
	
	\begin{figure*}[ht!]
		\centering
		\includegraphics[width=\linewidth]{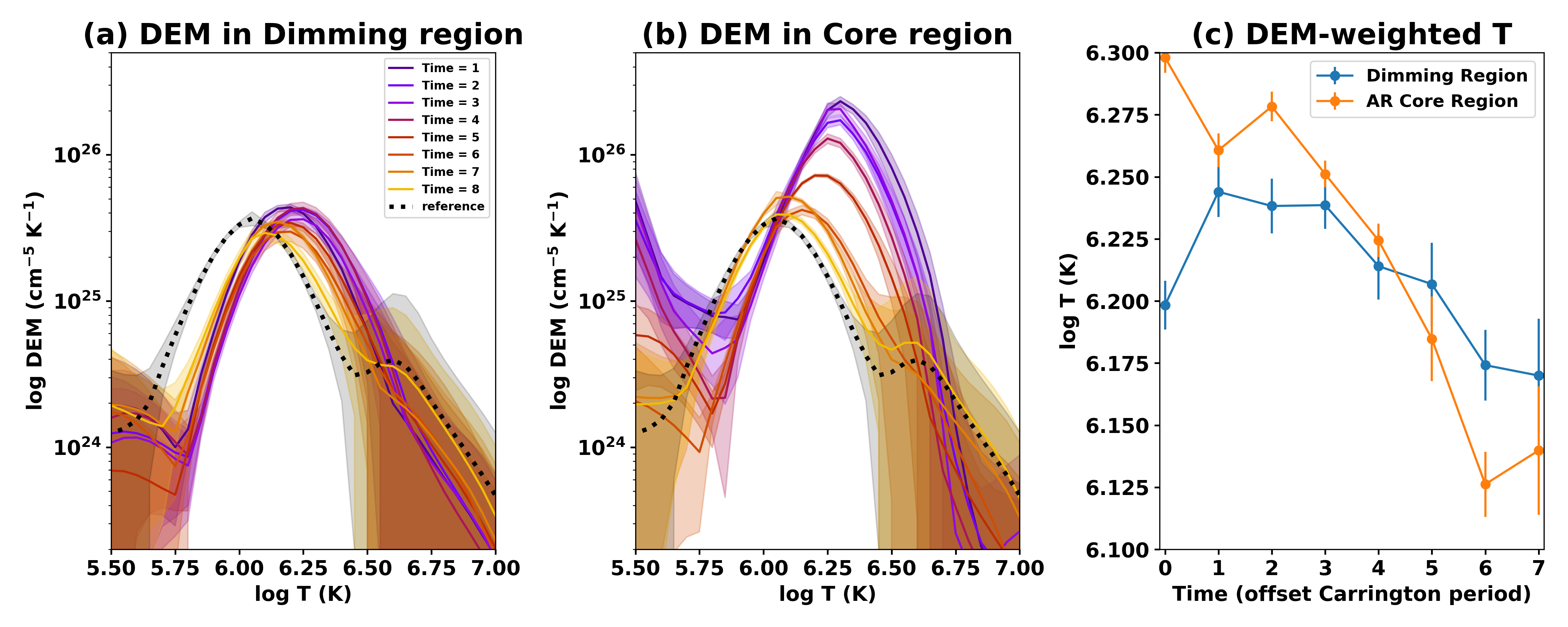}
		\caption{The DEM curves (a) in the dimming region and (b) in the AR centering region. The grey solid lines in panel (a) and (b) are the time-averaged DEM curves in the selected quiescent region (refers to the text). DEM uncertainties are shown as colored areas. Panel (c) shows the DEM-weighted temperature varying with time in the two sets of regions. }
		\label{fig:f6}
	\end{figure*}
	
	The DEM curves of the dimming region and the AR centering region at different times are shown in Figure~\ref{fig:f6}. The broadening in the DEM curves can be found both in the dimming region and the AR core region, which suggests the fading of the intense heating and meantime the appearance of cooling. In the dimming region, the EM around T = 10$^{5.7}$~K shows a significant increase after time 3, which indicates the recovery of the dimming region after the formation of the filament channel. We also chose a quiescent region outside the target AR to calculate the DEM-T distribution and found that its DEM curves barely vary with time. The time-averaged DEM curve of the quiescent region is shown by the grey solid line in Figure~\ref{fig:f6}(a-b). The dimming region has a significant lack of EM around T = 10$^{5.7}$K. In the AR core region, the EMs in the higher temperature (i.e. higher than T = 10$^{6.0}$~K) exceed the quiescent reference much enough to compensate the shortage of EM around T = 10$^{5.7}$~K, considering that the response function of AIA 171 \AA\ waveband has a wide temperature spread, so that the AR core region on the AIA 171 \AA\ images does not show a dark characteristic. Figure~\ref{fig:f6}c is the DEM-weighted temperature calculated according to Eq. (2). The integration range of temperature is set to log(T)=5.7 to 7.1, which include temperature in quiet Sun and in active region when there is no significant flare event. The temperature maintains a relatively steady state in the dimming region for the first four periods and drops down after the existence of the filament channel, and returns to a steady state in the last 3 periods. The temperature in the AR core region shows a general decreasing trend which agrees with the decay of the whole AR.

	\subsection{Magnetic Field Analysis}\label{sec:4.2}
	
	\begin{figure*}[ht]
		\centering
		\includegraphics[width=\linewidth]{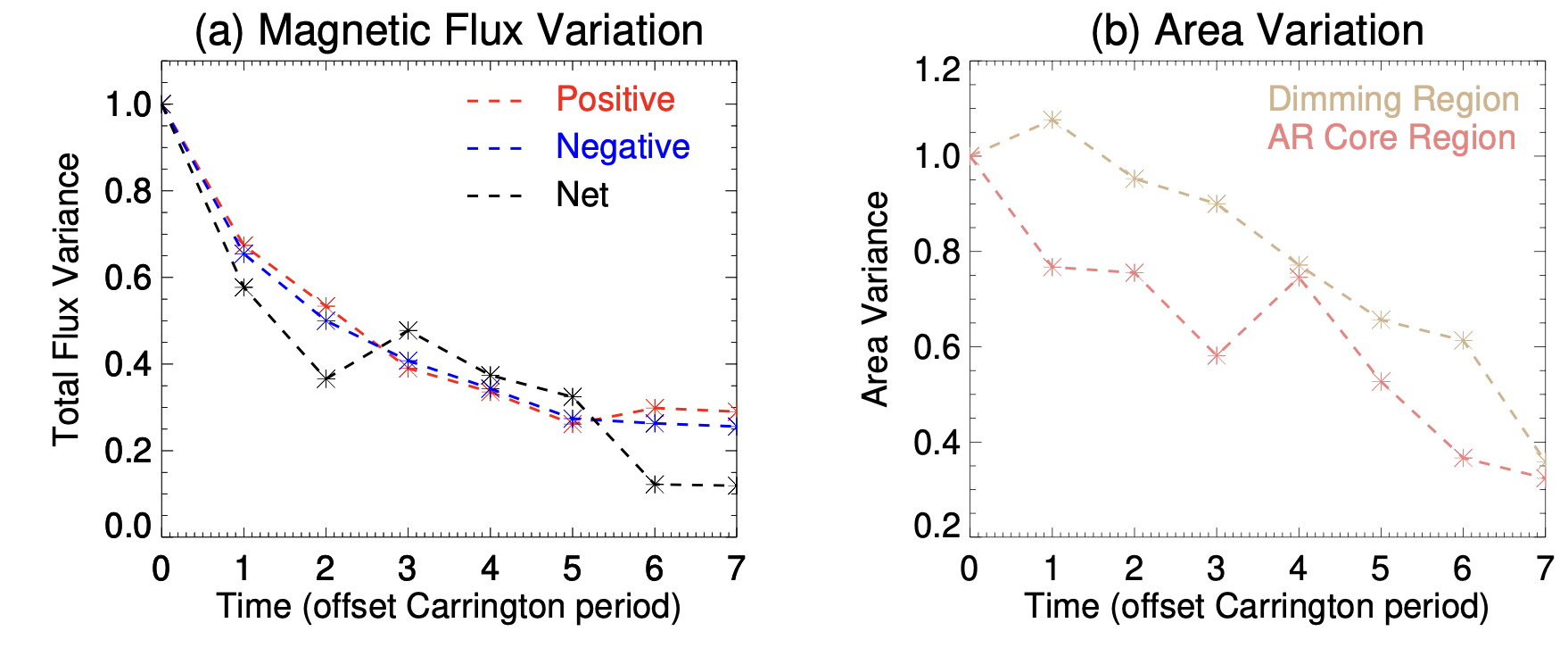}
		\caption{Variation of magnetic flux and area. (a) shows the average magnetic flux variation. Red/blue/black represent the positive/negative/net flux. (b) shows the area variation in the two sets of regions. Brown and red curves represent for the dimming region and the AR core region, respectively. Stars are the relative flux calculated from observation, while the dashed lines show the varying trend. }
		\label{fig:f5}
	\end{figure*}
	
	Figure~\ref{fig:f5}a displays the variation of the magnetic flux. The magnitude of both the positive and negative magnetic flux has decreased since the first detection of the AR, and this trend slows down after the existence of a filament channel (between the offset Carrington period 3 and 4). The continual decrease of the net flux yields the decay of the magnetic activity.
	Figure~\ref{fig:f5}b shows a general descending trend of areas in both the dimming region and the AR core region since Apr. 14. Also, the decreasing speed in the AR core region steepened after the formation of the filament channel at time 4. The anomaly decrease of the AR core area at time 3 may be due to the small coronal hole penetrating the central part of the AR. The area of the dimming region shows a roughly constant speed of decreasing.
	
	\begin{figure*}[ht]
		\centering
		\includegraphics[width=\textwidth]{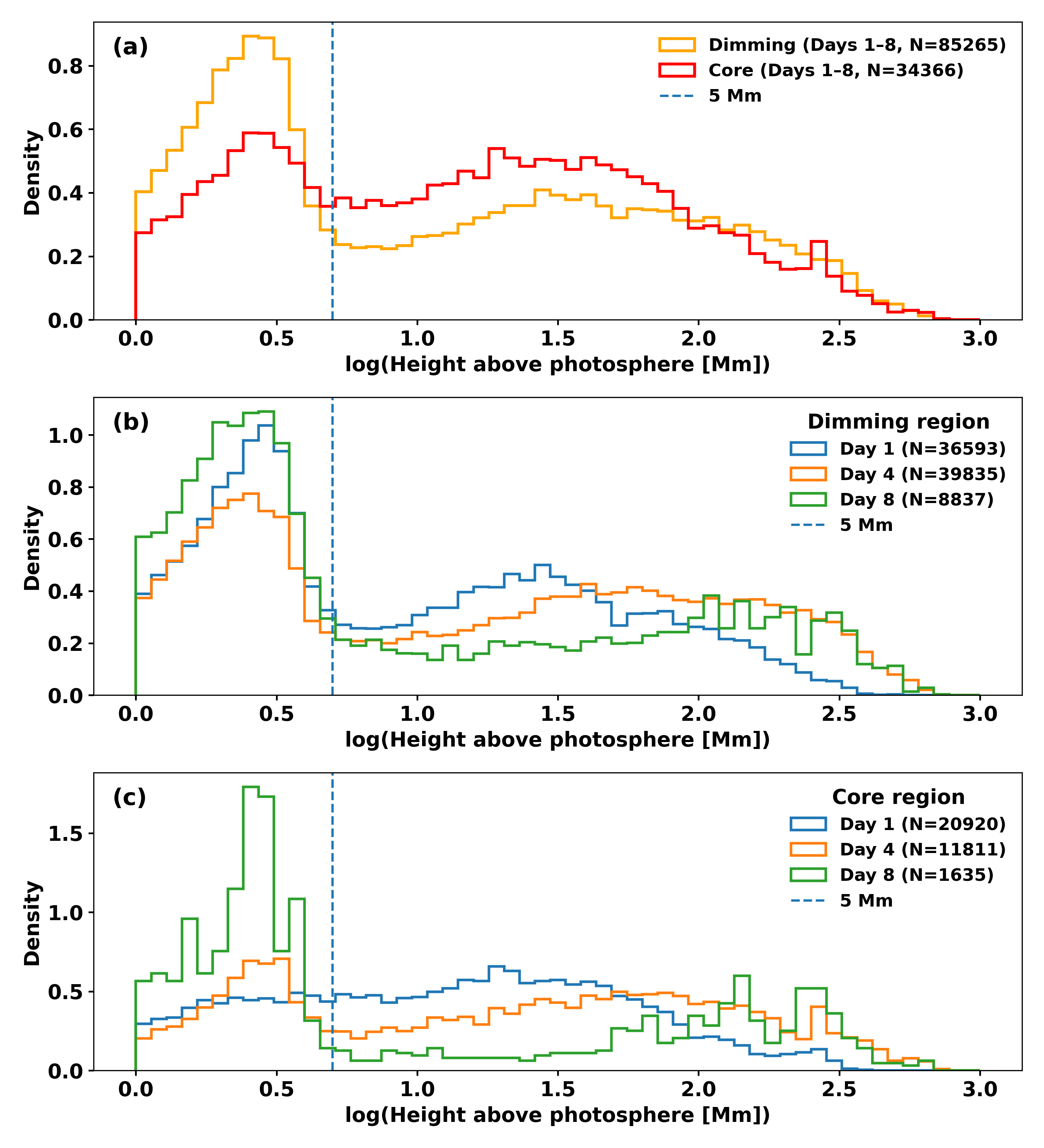}
		\caption{The average heights of the loops in the dimming region and the AR core. (a) shows distributions of time-averaged loop heights in dimming (orange) and AR core(red). (b) shows distributions of loop heights in dimming reigon at Time 1, 4 and 8. (c) shows distributions of loop heights in AR core. The dashed vertical line indicates the height of 5 Mm.}
		\label{fig:f7}
	\end{figure*}
	
	For further understanding of the magnetic field structures in the dimming region and the AR core region, we traced the field lines in the two sets of regions in the extrapolated PFSS magnetic fields. The distribution of the area- and time-averaged loop heights in the dimming region and the AR core are plotted in Figure~\ref{fig:f7}a. 
	It is noticeable that the dimming region has more loops form with height $\sim$1.5--5~Mm (log height = [0.2,0.7]), while the core region has more loops with height higher than 20~Mm (log height = 1.3), which indicates that the loops in dimming region are basically lower than those in the AR core.
	Distributions of loops heights at Carrington Time 1, 4, and 8 are presented to show temporal change in dimming region (Figure~\ref{fig:f7}b) and AR core (Figure~\ref{fig:f7}c). The dimming region shows two peaks in $\sim$3.5~Mm and above 20~Mm during the AR evolution. While the AR core shows a peak of distribution in $\sim$20 and 65~Mm at Time 1 and 4. At Time 8, the distributions of loop heights in both regions show similar trend. 
	
	It is important to emphasize that the height of coronal loops is not itself the cause of their temperature. Rather, both height and temperature are consequences of the magnetic field configuration and heating processes. Loops rooted in strong-field core regions typically receive more intense heating, reaching higher temperatures ($>$10$^{6.0}$ K) and, due to the expansion of magnetic flux with height, often extend to greater altitudes. Conversely, loops in peripheral regions, rooted in weaker fields, receive less heating and remain cooler. 
	The openness of fields is evaluated at R = 1.5$R_\mathrm{sun}$, where most of the field loops in the core region are closed and only south-west part of the dimming region has persistent positive open flux, which decay at a similar rate to magnetic flux.

	\section{Discussion}\label{sec:5}
	
	The dimming region is characterized by a two-component loop structure: low-lying loops with heights below 5 Mm and high-lying loops with temperatures exceeding the main response temperature of the 171 \AA\ waveband. According to \cite{1986ApJ...301..440A}, loops below 5 Mm typically reach temperatures under 10$^{5.5}$ K, which lies below the 171 \AA\ response range. 
	In the dimming region, PFSS extrapolation reveals high-lying loops that are magnetically connected to the AR core (Figure~\ref{fig:f3}). DEM analysis of the core region shows abundant plasma at temperatures $>$10$^{6.0}$ K (Figure~\ref{fig:f6}b), indicating sustained heating. We therefore infer that these connected high-lying loops in the dimming region are also heated to temperatures exceeding the 171 \AA\ response range (peak at 10$^{5.8}$ K). These hot loops overlap with cooler, low-lying loops ($<$10$^{5.5}$ K) that also contribute little to 171 \AA\ emission. The resulting absence of plasma at the specific temperature where 171 \AA\ is most sensitive produces the observed dimming instead of absence of plasma altogether.
	In contrast, the AR core is dominated by high-lying loops that are heated to temperatures matching the main response of the 193 \AA\ and 211 \AA\ wavebands. The significant excess of emission measure (EM) at these higher temperatures compensates for the minor deficit near the 171 \AA\ response, preventing dimming signatures in the core at this wavelength.
	
	AR 12738 was first observed on 14 April 2019, already formed an AR, precluding analysis of its emergence phase. One Carrington rotation later, a new AR emerged to its west but dissipated before Time 4. The interaction between these two regions may account for the temporary increase in 211 \AA\ intensity observed in the dimming region. 
	In the AR core, the DEM-weighted temperature declined steadily due to the loss of high-temperature plasma (e.g., at 193 \AA\ and 211 \AA\ temperatures). In the dimming region, cooling increased the EM around 10$^{5.7}$ K while reducing it near 10$^{6.3}$ K, resulting in a stable DEM-weighted temperature of approximately 10$^{6.2}$ K. Magnetic diffusion, ongoing since April 14, nearly ceased following the filament channel's deformation. As the dimming gradually recovered and the sunspot fragments disappeared, the AR eventually merged into the surrounding quiet Sun.
	An important caveat concerns the role of the filament channel that appears at Time 4. While its formation temporally coincides with the onset of cooling and dimming recovery, we cannot establish a causal relationship with the present data. The filament channel may actively participate in cooling overlying loops through thermal contact or magnetic reconnection, or it may simply be another manifestation of the same underlying process -- the continued decay and simplification of the AR's magnetic field. Distinguishing between these possibilities would require tracking the thermal evolution of individual loops throughout the filament eruption, ideally with spectroscopic diagnostics that can measure flow velocities and thermal structure along the line of sight. We therefore present the filament channel as an observational marker of the transition phase, while leaving its physical role to future investigation.
	
	Our findings add a new perspective to the study of coronal dimming by documenting a rare case of long-duration, non-eruptive dimming associated with the decay of AR 12738. Unlike impulsive dimmings linked to CME-driven magnetic reconnection \citep{2022ApJ...928..154J}, this dimming evolves gradually over several months, with DEM diagnostics revealing persistent cooling (log T $\approx$ 5.8). The dual magnetic loop configuration inferred from PFSS extrapolation suggests a stable restructuring process. Notably, the onset of filament channel formation appears to mark the transition to thermal recovery, pointing to filament buildup as a potential endpoint of slow dimming evolution.
	
	\section{Summary}\label{sec:6}
	
	This study presents the first systematic analysis of a long-lived peripheral dimming region, tracking its six-month evolution in relation to the decay of AR 12738. Unlike previous studies focused on short-term dimmings or radiance--flux relationships \citep[][]{2015ApJ...815...90U, 2017ApJ...846..165U}, we reveal the coupled thermal and magnetic evolution of a persistent dimming region. A key finding is a persistent DEM deficit in the 10$^{5.5}$--10$^{5.9}$ K range.
	This finding partially aligns with the mechanism proposed by \cite{2021ApJ...909...57S}, who suggested that strong magnetic pressure from overlying fields can suppress loops to low altitudes where temperatures remain below the 171 \AA\ response range. Such suppression could explain the presence of cool plasma ($<$10$^{5.5}$ K) in the dimming region. However, this mechanism alone cannot account for the simultaneous presence of hot plasma ($>$10$^{6.0}$ K) from high-lying loops connected to the core, nor for the specific DEM deficit centered precisely at 10$^{5.8}$ K. The magnetic pressure suppression would predict uniformly cool plasma throughout the dimming region, whereas our observations show a bimodal temperature distribution -- plasma that is either too cool or too hot for 171 \AA\ emission, with a deficit specifically at the intermediate temperature where this passband is most sensitive. The dominant dimming mechanism is therefore thermal restructuring of high-lying loops, not simply altitude suppression.
	
	PFSS extrapolation further reveals a structural dichotomy: the dimming region is dominated by low-lying loops ($<$ 5 Mm), while the AR core contains predominantly higher loops. This difference underpins their distinct DEM and emission profiles. During the late decay phase, the formation of a filament channel likely is followed by cooling of the high-lying loops, a process consistent with the heating-to-cooling transition \citep{2021ApJ...912....1P}, and observed here for the first time in the recovery of a long-term dimming region.
	
	Throughout its evolution, AR 12738 exhibited magnetic diffusion, filament channel formation and deformation, loop cooling, and dimming recovery. The area and average intensity in the 193~\AA\ and 211~\AA\ wavebands decreased as the region decayed, accompanied by a decline in mean magnetic field strength -- a trend that slowed following filament channel formation. The peripheral dimming, widespread in the 171~\AA\ channel during the early decay phase, resulted from a lack of plasma at the corresponding temperature due to the presence of low-lying loops below 10$^{5.5}$~K and high-lying loops heated beyond the 171~\AA\ response range. As the decay progressed, cooling loops began to cover the low-lying structures, increasing the 171~\AA\ intensity, after which dimming recovered. Ultimately, the AR's magnetic field became incorporated into the surrounding quiet Sun. The differing loop height distributions between the dimming and core regions, rooted in distinct magnetic morphologies, account for the observed differences in emission and DEM properties.

	\normalem
	\begin{acknowledgements}
		
		This work was supported by the National Key R\&D Program of China No. 2022YFF0503800, 2021YFA0718600, NSFC grant 12303057, Beijing Natural Science Foundation (grant NO. 1254055), China's Space Origins Exploration Program, and Specialized Research Fund for State Key Laboratory of Solar Activity and Space Weather.
		We appreciate the helpful discussion with Dr. Xudong Sun (University of Hawaii) and Dr. Hui Tian (Peking University). 
		We also thank the anonymous reviewer for his/her constructive comments and suggestions.
		We acknowledge the data use from SDO. SDO is a mission of NASA's Living With a Star Program.

	\end{acknowledgements}
	
	\bibliographystyle{raa}
	\bibliography{reference}
	
\end{document}